\documentclass[aps,prb,preprint,groupedaddress]{revtex4-1}
\usepackage{graphicx}
\usepackage{amsmath}
\usepackage{amsfonts}

\begin{document}
\newcommand{\guido}[1]{[\textbf{GUIDO:} \emph{#1}]}
\newcommand{\miquel}[1]{[\textbf{MIQUEL:} \emph{#1}]}

\title{Symmetries in the collective excitations of an electron gas in core-shell nanowires}

\author{Miquel Royo}
\email{miguel.royovalls@unimore.it}
\affiliation{Departament de Qu\'{\i}mica F\'{\i}sica i Anal\'{\i}tica, Universitat Jaume I, E-12080, Castell\'o, Spain}
\affiliation{CNR-NANO S3, Institute for Nanoscience, Via Campi 213/a, 41125 Modena, Italy}
\author{Andrea Bertoni}
\email{andrea.bertoni@nano.cnr.it}
\affiliation{CNR-NANO S3, Institute for Nanoscience, Via Campi 213/a, 41125 Modena, Italy}
\author{Guido Goldoni}
\email{guido.goldoni@unimore.it}
\affiliation{Department of Physics, Informatics and Mathematics, University of Modena and Reggio Emilia, Italy}
\affiliation{CNR-NANO S3, Institute for Nanoscience, Via Campi 213/a, 41125 Modena, Italy}

\date{\today}

\begin{abstract}
We study the collective excitations and inelastic light scattering cross-section of an electron gas confined in a GaAs/AlGaAs coaxial quantum well. These system can be engineered in a core-multi-shell nanowire and inherit the hexagonal symmetry of the underlying nanowire substrate. As a result, the electron gas forms both quasi 1D channels and quasi 2D channels at the quantum well bents and facets, respectively. Calculations are performed within the RPA and TDDFT approaches. We derive symmetry arguments which allow to enumerate and classify charge and spin excitations and determine whether excitations may survive to Landau damping. We also derive inelastic light scattering selection rules for different scattering geometries. Computational issues stemming from the need to use a symmetry compliant grid are also investigated systematically.
\end{abstract}


\maketitle


\section{Introduction}

Semiconductor nanowire (NW) lateral heterostructures -- such as coaxial heterojunctions or quantum wells -- represent an important new
class of nanomaterials with promising versatile properties for future applications in nanotechnology.~\cite{LieberMrsB07} Many of
their advantages derive from the high precision and reproducibility of ``bottom-up'' NW growing techniques,
~\cite{HyunARMR13} which allows for near-ideal atomically sharp interfaces to be engineered both in the axial and in the radial
direction. Therefore, heterostrured NWs provide the possibility of tuning quantum confinement properties by band structure engineering
in the radial direction, while using the extended axis for facile transport and device integration, including with Si substrates
thanks to the strain release in mismatched NW interfaces.~\cite{KastnerPM04,MartenssonNL04,GlasPRB06,RoestNT06,TomiokaSTQE11}

Coaxial heterostructures may also host a high-mobility electron gas (EG).~\cite{FunkNL13} The remote doping technique has been recently demonstrated in $\mathrm{GaAs/Al_{x}Ga_{1-x}As}$ core-shell NWs, with the EG confined at the NW heterojunctions.~\cite{SpirkoskaPSSRRL11,BertoniPRB2011,WongNL2011} Hole gas in unintentionally doped structures can also be realized.~\cite{Paulina2014} This opens up the realization of a variety of NW-based electronic devices.~\cite{LuTED08,LiNL06,TomiokaNAT12}. On the other hand, this allows the investigation of fundamental properties of the EG in novel nanoscopic morphologies.~\cite{BertoniPRB2011,WongNL2011}

Traditional probes of the EG based on transport measurements, such as the Hall mobility, are difficult in NWs due to the difficulty of creating the required Ohmic contacts.~\cite{PersanoNT12,WirthsAPL12} On the other hand, optical spectroscopies are nondestructive and contactless probes. Dynamics of photoexcited electron-hole plasmas were studied by PL in single bare NWs~\cite{TitovaNL07} and core-shell NWs.~\cite{HoangNL07} More recently, pump-THz probe spectroscopy allowed to determine mobilities, lifetimes and surface recombination rates of photoexcited carriers in different III-V NWs~\cite{JoyceNT13} and core-shell NWs.~\cite{IbanesAPL13} Inelastic light scattering (ILS) has been used to extract carrier density and mobility data from the plasmon-phonon coupling modes of photoexcited NWs~\cite{KettererNS12} and multilayered NWs.~\cite{KettererPRB11} Recently, we have used mean-field simulations combined with ILS experiments to demonstrate that remote-doping induces high-mobility EG, and to assign ILS resonances to separate quasi-1D (q1D) and quasi-2D (q2D) channels in the sample. ~\cite{FunkNL13}

Indeed, ILS has been used for many years to study collective excitations of excess carriers in semiconductor heterostructures,~\cite{ILScardonaBook,SchullerBook2006} as it enables to detect charge-density excitations (CDE) and spin-density excitations (SDEs) separately, and, under strong resonant conditions, unscreened single-particle excitations (SPEs).~\cite{BlumPRB70,KatayamaJPSJ85,HawrylakPRB85,DasSarmaPRL99,KushwahaAIPA12,SteinebachPRB96,TavaresPRB05} Comparison of ILS with theoretical models allows to obtain subband structure, electron density, mobilities, many-body interactions, etc. In fact, collective excitations are strongly dependent on dimensionality, and different modeling apply for, e.g., planar quantum wells (QWs),~\cite{BlumPRB70,KatayamaJPSJ85,HawrylakPRB85,DasSarmaPRL99, KushwahaAIPA12} quantum wires~\cite{SteinebachPRB96,TavaresPRB05,DasSarmaPRB96}, or quantum dots~\cite{Garcia2005,Kalliakos2008}. Collective excitations of an axial EG have been investigated theoretically in cylindrical geometries~\cite{SatoPRB93,WendlerPRB94,WangPRA02} which somehow iterpolates between 1D and 2D, but neglecting any discrete symmetry of realistic devices. This is a particularly severe approximation, as in core-shell NWs, be they doped heterojunctions or QWs, the cylindrical symmetry is relaxed as a result of the prismatic form of the substrate NW. As a consequence, q1D and q2D channels are invariably formed in the sample.\cite{FerrariNL08,BertoniPRB2011,FunkNL13}

In this paper we study the collective excitations of an EG confined in a hexagonal coaxial QW (coQW), as engineered in a core-multishell $\mathrm{GaAs/Al_{x}Ga_{1-x}As}$ NW. Channels showing q1D and q2D character are subsequently populated by varying the Fermi energy. We adapt the multi-subband RPA and TDLDA formalisms, well established from studies in lower dimensional systems, to perform a full 3D modeling of the NW electronic excitations, tracing the calculated CDEs and SDEs to the hexagonal symmetry of the system. Group theory is used to classify the complex set of collective excitations and it is shown that Landau damping into single-particle excitations takes place only for excitations of matching symmetry. Finally, we obtain ILS cross-sections and predict ILS spectra under different scattering geometries, showing that the anisotropy of the system may be clearly exposed in ILS spectroscopy. The need for a proper calculations of single-particle states and Coulomb matrix elements in a symmetry compliant grid are emphasized.

The article is organized as follows. In Sec.~\ref{theory}, we sketch the theoretical model. We use the TDDFT formalism with one invariant direction to describe the coQW system (\ref{theo_TDDFT}) and we formulate the non-resonant formalism employed to compute the ILS spectra (\ref{theo_ILS}). Finally, in subsection~\ref{theo_details} the details of the computational procedure are summarized.
In Sec.~\ref{Results} we illustrate CDE and SDEs at various carrier densities. In subsection~\ref{sec:SingleParticleStates} we describe the single-particle states used as basis set
to compute the excitations. In subsection~\ref{Res_elemexc} we report our RPA and TDDFT results and, finally, in subsection~\ref{ILS} we report ILS spectra computed for two relevant scattering geometries.
To conclude, in Sec.~\ref{concl} we summarize and discuss our results. In Appendix \ref{coulomb_app} we discus the calculation of Coulomb integrals on a triangular grid.

\section{Theoretical model \label{theory}}

\subsection{The linear-response TDDFT  approach for q1D systems \label{theo_TDDFT}}

In linear-response theory the excitation energies of an interacting electron system can be obtained
from the poles of the density-density response function.~\cite{giulianiBook} In the Lehman representation this so-called \emph{irreducible response function} is written as,

\begin{equation}
\tilde{\Pi}(\mathbf{R},\mathbf{R}',\omega)= \sum_{n=1}^{\infty} \left\{ \frac{\langle\Psi_0\lvert
\hat{n}(\mathbf{R}) \rvert \Psi_n \rangle \langle\Psi_n\lvert \hat{n}(\mathbf{R}') \rvert \Psi_0 \rangle}
{\omega -\Omega_n + i\eta} -  \frac{\langle\Psi_0\lvert
\hat{n}(\mathbf{R}') \rvert \Psi_n \rangle \langle\Psi_n\lvert \hat{n}(\mathbf{R}) \rvert \Psi_0 \rangle}
{\omega +\Omega_n + i\eta} \right\}.
\label{eq1}
\end{equation}

\noindent Here, $\Psi_0$ and $\Psi_n$ are the many-electron ground and excited states wave-functions,
respectively, $\Omega_n=E_n-E_0$ are the excitations energies, $\hat{n}(\mathbf{R})$ is the density operator
expressed in the spatial coordinate $\mathbf{R}$, and $\eta$ is a positive small damping parameter.
In the TDDFT formalism~\cite{ullrichBook} $\tilde{\Pi}(\mathbf{R},\mathbf{R}',\omega)$ is obtained from the response function
of the non-interacting Kohn-Sham (KS) system $\Pi^0(\mathbf{R},\mathbf{R}',\omega)$. The latter is formally
obtained from Eq.~(\ref{eq1}) evaluating the matrix elements in the numerator assuming single Slater
determinants built from the KS orbitals. For a NW translationally invariant along the $z$ direction, 
the latter can be factorized as,

\begin{equation}\label{eq:factorization}
    \varphi_{n\,k_z}(\mathbf{R})=\phi_n(\mathbf{r}) e^{i k_z z},
\end{equation}

\noindent where $\phi_n(\mathbf{r})$ is an envelope function over the NW in-plane directions $\mathbf{r}\equiv(x,y)$ and $k_z$ is the momentum along the in-wire direction $z$. Correspondingly, the energy of the KS states $\varepsilon_{n}(k_z)$ is parabolic in $k_z$. Likewise, the density operator is conveniently Fourier transformed along the invariant
direction which yields, $\hat{n}(\mathbf{r},q_z)=\sum_{l=1}^N \delta(\mathbf{r}-\mathbf{r}_l) e^{-iq_zz_l}$,
$N$ being the total number of electrons. Altogether, the KS response function reads,

\begin{equation}
\Pi^0(\mathbf{r},\mathbf{r}',q_z,\omega)= \sum_{ij} \Pi_{ij}^0(q_z,\omega)\phi_i^*(\mathbf{r}) \phi_j(\mathbf{r})
\phi_i(\mathbf{r}') \phi_j^*(\mathbf{r}'),
\label{eq2}
\end{equation}

\noindent with

\begin{equation}
\Pi_{ij}^0(q_z,\omega)= g \int \frac{dk_z}{2\pi} \frac{f_i(k_z)-f_j(k_z+q_z)}{\omega-(\varepsilon_j(k_z+q_z)
-\varepsilon_i(k_z))+i\eta},
\label{eq3}
\end{equation}

\noindent where $g=2$ accounts for electron spin degeneracy, $f_n(k_z)$ is the Fermi occupation function,
and $q_z$ is the change in the in-wire momentum.

To obtain the response function of the interacting system we expand it in terms of the KS orbitals as follows,

\begin{equation}
\tilde{\Pi}(\mathbf{r},\mathbf{r}',q_z,\omega)= \sum_{ijlm} \tilde{\Pi}_{ijlm}(q_z,\omega)\phi_i^*(\mathbf{r}) \phi_j(\mathbf{r})
\phi_l(\mathbf{r}') \phi_m^*(\mathbf{r}').
\label{eq4}
\end{equation}

\noindent The matrix elements $\tilde{\Pi}_{ijlm}(q_z,\omega)$ are then related to the elements of the KS response function
(\ref{eq3}) through the following Dyson equation,

\begin{equation}
\sum_{ijlm}\tilde{\Pi}_{ijlm}(q_z,\omega)=
\sum_{ijlm}  \Pi^0_{ij}(q_z,\omega)\, \delta_{il}\,\delta_{jm} +
\sum_{ij} \Pi^0_{ij}(q_z,\omega)
\sum_{knlm}
\left[v_{ijkn}(q_z)+u^{XC}_{ijkn} \right]
\tilde{\Pi}_{knlm}(q_z,\omega).
\label{eq5}
\end{equation}

\noindent Here, $v_{ijkn}(q_z)$ and $u^{XC}_{ijkn}$ are the direct Coulomb and
exchange-correlation matrix elements, respectively, which describe the dynamic interaction of two electrons, one of
which gets scattered from state $i$ to $j$ and the other from $k$ to $n$, with an exchange of momentum $q_z$.

The direct Coulomb matrix elements read

\begin{equation}
v_{ijkn}(q_z)= \int d\boldsymbol{r}
\int d\boldsymbol{r}' \, \phi_i(\boldsymbol{r})\, \phi_j^*(\boldsymbol{r}) \,
\hat{V}(\boldsymbol{r}-\boldsymbol{r}',q_z)\,
\phi^*_k(\boldsymbol{r}')\, \phi_n(\boldsymbol{r}'),
\label{eq6}
\end{equation}

\noindent where $\hat{V}(\boldsymbol{r}-\boldsymbol{r}',q_z)$ is the Fourier transform of the Coulomb operator in the
$z$ direction.

The exchange-correlation matrix elements read

\begin{equation}
u^{XC}_{ijkn}= - \int d\boldsymbol{r}
 \, \phi_i(\boldsymbol{r})\, \phi_j^*(\boldsymbol{r}) \,
\hat{f}_{XC}(\boldsymbol{r})\,
\phi^*_k(\boldsymbol{r})\, \phi_n(\boldsymbol{r}),
\label{eq7}
\end{equation}

\noindent with $\hat{f}_{XC}$ being the dynamic exchange-correlation kernel defined in the adiabatic local
density approximation as the derivative of the static exchange-correlation potential
with respect to the ground state density, $\hat{f}_{XC}(\boldsymbol{r}) =\frac{dV^{XC}(\boldsymbol{r})}
{dn(\boldsymbol{r})}$. In particular, using the LDA functional conceived by Gunnarsson
and Lundqvist~\cite{gunnarssonPRB76} the exchange-correlation kernel reads~\cite{LuoPRB93}

\begin{equation}
\hat{f}_{XC}(\boldsymbol{r})=-1.704\,a_0^*(\mathbf{r})^3 r_s(\mathbf{r})^2 \left[ 1+\frac{0.6213\,r_s(\mathbf{r})}
{11.4+r_s(\mathbf{r})} \right] Ry^*(\mathbf{r}).
\label{eq8}
\end{equation}

\noindent Here, $a_0^*(\mathbf{r})$ and $Ry^*(\mathbf{r})$ are the material dependent effective Bohr radius and
Rydberg energy, respectively, and $r_s(\mathbf{r})$ is the Wigner-Seitz radius.

The imaginary part of the irreducible response function (\ref{eq4}) is proportional to the dynamic
structure factor according to the fluctuation-dissipation theorem. As such, it gives a direct measure of the spectral strength of various
elementary excitations of the electron system. In the long wavelength limit ($q_z\rightarrow 0$), collective charge density
excitations (CDEs) or plasmons carry all the spectral weight and show up as narrow spectral peaks.
However, as $q_z$ increases electron-hole single-particle excitations (SPEs) start gaining spectral
weight and show up as weak broad bands.

The TDDFT formalism also describes spin-density excitations (SDEs). The latter
are only coupled through indirect-exchange interactions. Consequently, they always appear redshifted
with respect to their plasmonic counterparts. The SDEs are obtained from the poles of the so-called \emph{reducible
response function}, $\Pi(\mathbf{R},\mathbf{R}',\omega)$. The latter is calculated within a procedure similar to the
above one setting to zero the direct Coulomb integrals in Eq.~(\ref{eq5}).
Furthermore, by deactivating the exchange-correlation integrals in Eq.~(\ref{eq5}) one recovers the CDEs
within the random phase approximation (RPA).

\subsection{ILS cross-section \label{theo_ILS}}

In a non-resonant formalism the ILS cross-section is estimated from the imaginary part of the appropriate complete momentum dependent response functions.~\cite{KushwahaAIPA12} For CDEs, we Fourier transform the irreducible response function to obtain

\begin{equation}
\tilde{\Pi}(\mathbf{Q}, \omega)= \sum_{ijlm} \tilde{\Pi}_{ijlm}(q_z,\omega)
\iint d\mathbf{r} \, d\mathbf{r}'
e^{-i\,\mathbf{q}(\mathbf{r}-\mathbf{r}')}
\phi^*_i(\mathbf{r})\, \phi_j(\mathbf{r})\,\phi_l(\mathbf{r}')\,\phi^*_m(\mathbf{r}').
\label{eq9}
\end{equation}

\noindent Here, $\mathbf{q}$ is the in-plane component of the total momentum $\mathbf{Q}\equiv(\mathbf{q},q_Z)$
exchanged in the scattering process, i.e., $\mathbf{Q}=\mathbf{Q}_i-\mathbf{Q}_s$, $\mathbf{Q}_i$ and $\mathbf{Q}_s$ being the momenta of the incident and scattered photons, respectively.

From the response functions we may also calculate the density fluctuation induced by the electromagnetic field at a given
energy and momentum, the so-called induced density distribution (IDD) from Kubo's correlation formula~\cite{KuboJPSJ57}

\begin{equation}
\delta n(\mathbf{r},\mathbf{q},q_z,\omega)= \int d\mathbf{r}' \, \tilde{\Pi}(\boldsymbol{r},\boldsymbol{r}',q_z,\omega)
\, e^{i\, \mathbf{q} \mathbf{r}'}.
\label{eq10}
\end{equation}

Scattering cross-section and IDD for SDEs are obtained analogously from the reducible response function, $\Pi(\mathbf{R},\mathbf{R}',\omega)$.

\subsection{Computational methods \label{theo_details}}

To evaluate the system response functions we need the static properties of the system, that is the single-particle
energies and corresponing envelope functions of the q1D subbands, as well as the total electron density. At the mean-field level, these should be obtained from a self-consistent DFT calculation.~\cite{FunkNL13} However, since in this paper we aim at the dynamic properties of the EG, we clear up the calculation of the
static properties by neglecting mean-field effects. In other words, the energy subbands are only determined by the unscreened band-offset modulation of the coQW. Since single-particle states would be different within different screening schemes, using a common unscreened confinement allows us to expose the difference in the dynamic properties within different formalisms, namely, TDLDA  and RPA. On the other hand, mean-field effects would be dominated by band-offset confinement in
narrow QWs as the present one. Mean-field calculations of static properties would be obviously required to study doped heterojunctions.~\cite{BertoniPRB2011}

The envelope-functions and subband energies are obtained within a single-band effective mass approximation. Assuming in-wire  spatial invariance along $z$ and factorizing the envelope functions as in (Eq.~\ref{eq:factorization}), the 2D envelope functions $\phi_n(\mathbf{r})$ are given by

\begin{equation}
\left[-\frac{\hbar^2}{2}\nabla_\mathbf{r}
\left(\frac{1}{m^*(\mathbf{r})} \nabla_\mathbf{r} \right)
+ V(\mathbf{r}) \right] \phi_n(\mathbf{r}) =
\varepsilon_n \phi_n(\mathbf{r}) \, ,
\label{eq11}
\end{equation}

\noindent where $V(\mathbf{r})$ is the spatial confinement potential determined by the core-shell band offset. Eq.~(\ref{eq11}) is numerically integrated in a hexagonal domain delimited by the NW surface using a symmetry-compliant triangular 
grid with $27.55~\mathrm{points/nm^2}$, and assuming Dirichlet boundary conditions.

In the following section we shall investigate the dynamical properties of the EG at different densities and subband occupations. This is performed by fixing the position of the Fermi energy to our interest, and then calculating the Fermi occupation of the subbands
(we assume zero temperature throughout this paper) and the electron density which is used in the dynamic exchange-correlation kernel, Eq.~(\ref{eq8}).

The computation of the Coulomb matrix elements entering the Dyson equation (\ref{eq5}) is the numerically most
intensive part of the procedure, as it requires to calculate the 4D integrals (\ref{eq6}) in a
dense grid. To speed up the calculation we have adapted the Fourier convolution theorem, which has been widely used
to calculate Coulomb integrals in rectangular grids,~\cite{WenigerPRA86} to the case of a triangular grid. The method is
described in Appendix~\ref{coulomb_app}. We also take advantage of the system symmetries to reduce the number of
Coulomb integrals to be computed as follows:
\begin{itemize}
  \item since the electron wave-functions $\phi_n(\mathbf{r})$ are real,
the following symmetries hold: $v_{ijkn}=v_{ijnk}=v_{jikn}=v_{jink}=v_{knij}=v_{knji}=v_{nkij}=
v_{nkji}$;
  \item as a consequence of the existence of a center of inversion in the system we
can classify the wave-functions as \emph{gerade} or \emph{ungerade}, which allows us to discard \emph{a priori} the computation of integrals with odd integrand;
  \item Coulomb elements in which the first two indexes ($ij$) correspond to empty subbands vanish. This is
because such elements are multiplied in the Dyson equation (\ref{eq5}) by the KS response function $\Pi_{ij}^0(q_z,\omega)$
which is zero for empty $ij$ subbands [see Eq.~(\ref{eq3})].
\end{itemize}
These very same considerations can be adopted
to reduce the calculation of the exchange-correlation terms. However, in that case the integrals are
easier to compute due to the locality of the LDA potential which reduces the dimensionality to 2D, and standard numerical integration methods can be used.

To compute the interacting response functions we rewrite the Dyson equation (\ref{eq5}) in tensor notation
as,

\begin{equation}
\mathbf{\tilde{\Pi}}=\mathbf{\Pi}^0\, \mathbf{I} + \mathbf{\Pi}^0 \left( \mathbf{v} +
\mathbf{u}^{\mathrm{XC}} \right) \mathbf{\tilde{\Pi}},
\label{eq12}
\end{equation}

\noindent or equivalently,

\begin{equation}
\mathbf{\tilde{\Pi}} \boldsymbol{\varepsilon} =\mathbf{\Pi}^0\,
\mathbf{I}.
\label{eq13}
\end{equation}

\noindent In the above equation, $\mathbf{I}$ is the identity matrix, and $\boldsymbol{\varepsilon} = \mathbf{I}
- \mathbf{\Pi}^0  \left(\mathbf{v} + \mathbf{u}^{\mathrm{XC}} \right)$ is the dielectric tensor of the electron
gas, whose inverse yields the required solution, $\mathbf{\tilde{\Pi}}=\mathbf{\Pi}^0\, \mathbf{I} \,
\boldsymbol{\varepsilon}^{-1}$. Recovering the subband notation leads to,

\begin{equation}
\tilde{\Pi}_{ijlm}(q_z,\omega)= \Pi_{ij}^0(q_z,\omega) \varepsilon^{-1}_{ijlm}(q_z,\omega).
\label{eq14}
\end{equation}

\noindent Therefore, in order to calculate a single element of the response function, we first build up
the complete dielectric tensor of the EG and then invert it by means of efficient routines.~\cite{MKL}

\section{Numerical Results \label{Results}}

\subsection{Single-particle states\label{sec:SingleParticleStates}}

The reference system for the following sections is a core-multishell NW (CSNW), as the one outlined in Fig.~\ref{fig1}(a), with an hexagonal $\mathrm{Al_{0.3}Ga_{0.7}As}$ central core 100 nm of diameter, a 20 nm wide shell of GaAs, and a 10 nm
$\mathrm{Al_{0.3}Ga_{0.7}As}$ capping layer. The 20 nm GaAs shell is a coQW for the conduction
electrons. Note that this is somehow different from usual samples, particularly in that the core is typically grown from GaAs. However, in doped samples the core is usually depopulated due to band bending.~\cite{BertoniPRB2011,FunkNL13} As we don't perform self-consistent calculations here, we use a $\mathrm{Al_{0.3}Ga_{0.7}As}$ core to exclude the formation of low-energy core states which we are not interested in.

In the calculation the GaAs/$\mathrm{Al_{0.3}
Ga_{0.7}As}$ band offset is taken as 0.284 eV~\cite{BosioPRB88} and the origin of energies is placed at the GaAs conduction
band edge. The position  dependent effective mass $m^*(\mathbf{r})$ is 0.067 in GaAs and 0.092 in $\mathrm{Al_{0.3} Ga_{0.7}As}$ regions.~\cite{M.Levinshtein96}

In Figs.~\ref{fig1}(b) and (c) we show the normalized DOS and the in-plane envelope functions $\phi_n(\mathbf{r})$,
respectively, for the lowest single-particle states. Two types of states can be identified, those preferentially localized in the corners and those which are delocalized over the coQW. The former tend to have lower energy.~\cite{FunkNL13} The envelope functions show an increasing number of nodal planes \emph{normal} to the coQW plane, which can be interpreted as the discretized momentum of a QW which is wrapped around the axis. States with nodal surfaces \emph{parallel} to the coQW plane, analogous to excited subbands in planar QWs, lie at higher energies and are not shown here.

For the discussion of collective excitations and their ILS cross-section, it will be useful to classify the single-particle states on the basis of their symmetry. Due to the overall hexagonal symmetry of the system, the eigenstates form basis of the irreducible representations of the $\mathrm{D_{6h}}$ group. In Fig.~\ref{fig1}(c) we label the symmetry irreducible representation (irrep) of each state, with E-type being doubly degenerate representations. The symmetry-induced degeneracies show up in the DOS as the higher peaks (see Fig.~\ref{fig1}(b)).

\subsection{Elementary excitations dispersion \label{Res_elemexc}}

\begin{figure}[h!]
\includegraphics[width=0.8\textwidth]{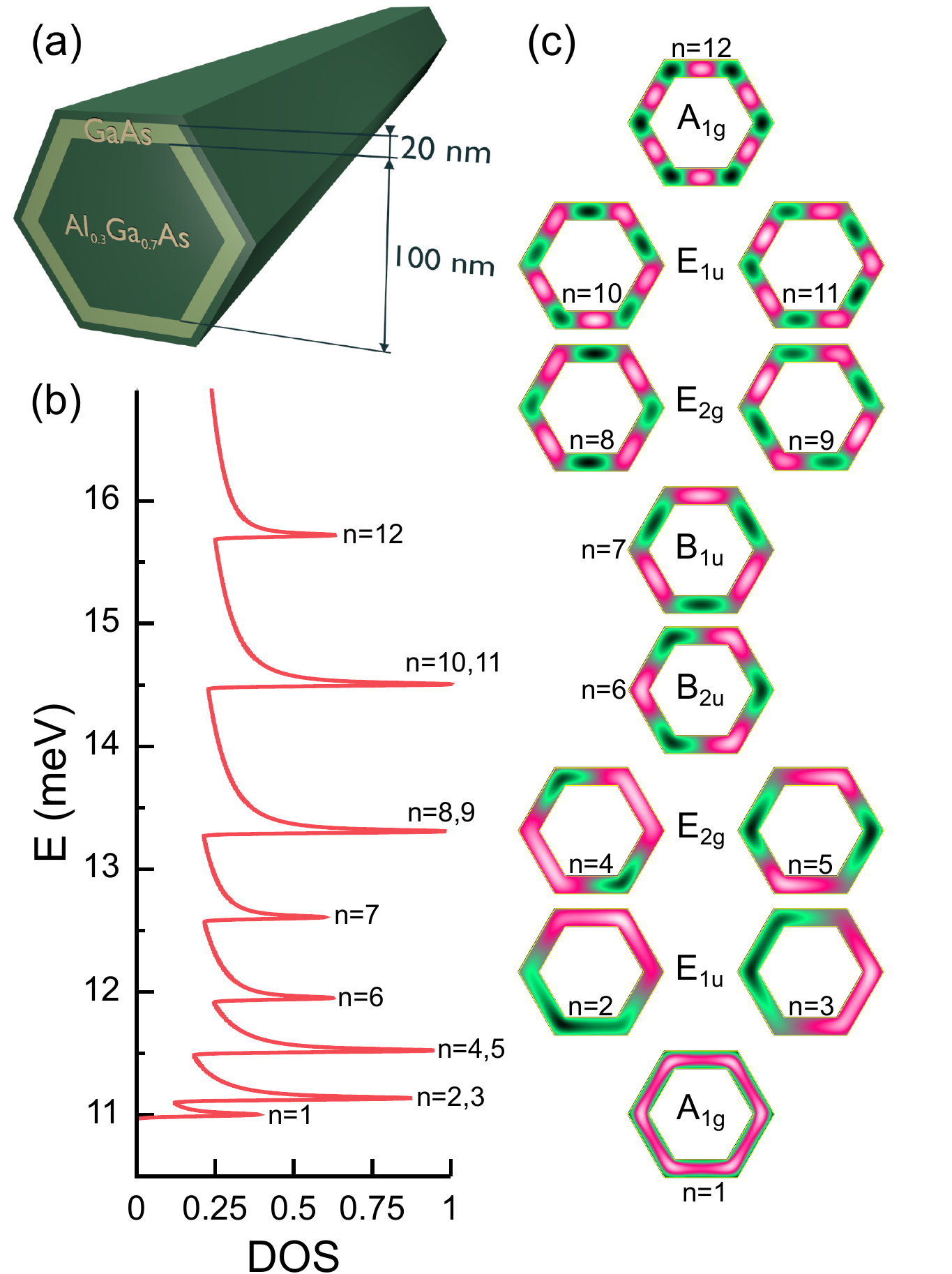}
\caption{(a) Schematics of the studied core-multishell NW, (b) normalized single-particle DOS in arbitrary units,
and (c) envelope functions of the 12 lowest lying subbands. The principal quantum number $n$ and the irrep 
of the $\mathrm{D_{6h}}$ group of each state is indicated.}
\label{fig1}
\end{figure}

\begin{figure}[h!]
\includegraphics[width=0.5\textwidth]{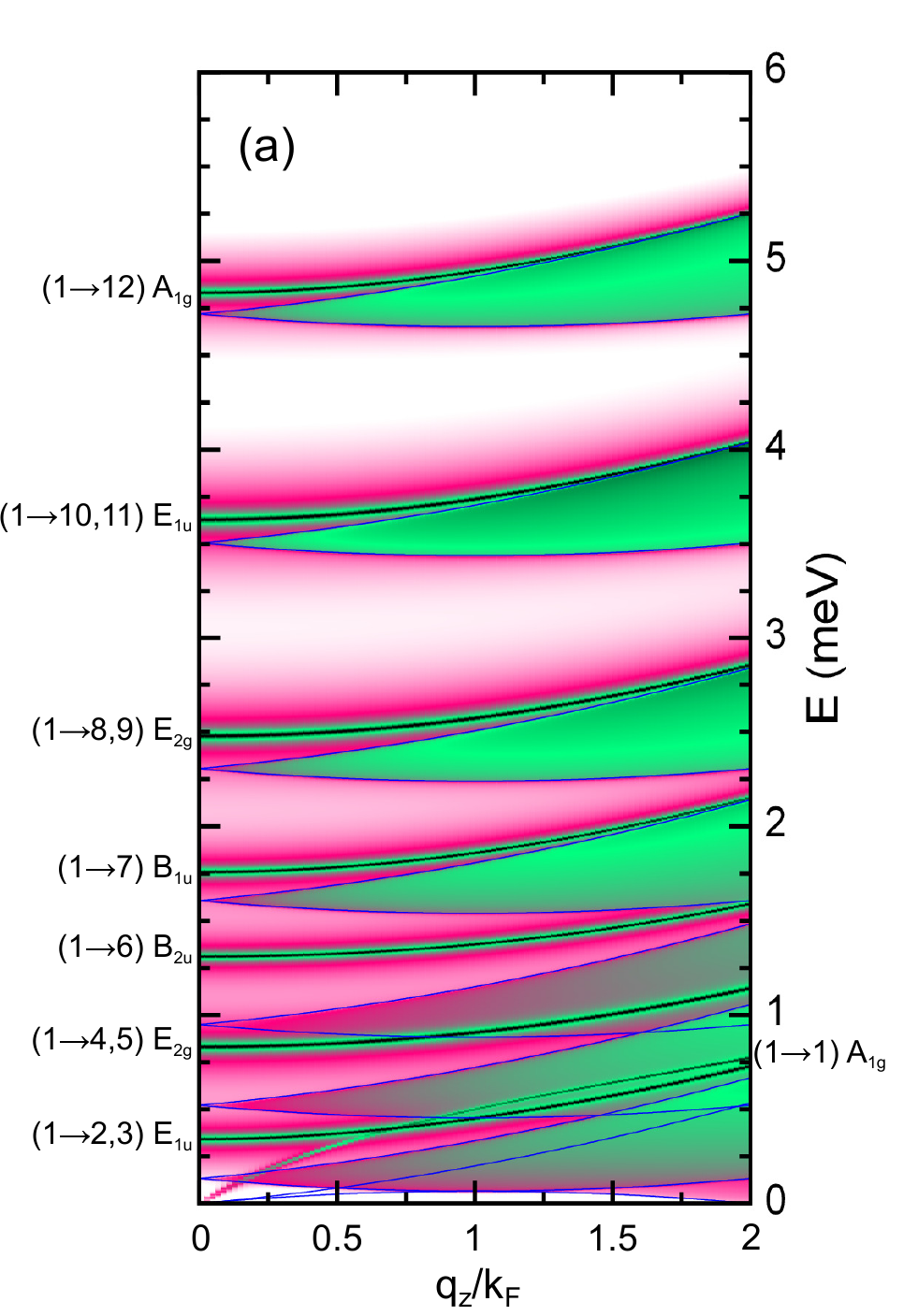}
\caption{(a)CDEs dispersion in $q_z/k_F$ for a NW with one occupied subband calculated within the RPA. The color map is plotted with
a logarithmic scale to emphasize the electron-hole single-particle excitation continua. The bounds of the latter are also
delimited by blue solid lines analytically calculated. The lateral labels show the subbands involved in the excitation and
the symmetry of the excited state.}
\label{fig2}
\end{figure}

We now look into the collective excitations of the EG in the coQW system at different density regimes, corresponding to occupation of an increasing number of subbands. Here we are interested in classifying the lowest energy excitations of the system, which correspond to CDE and SDE along the NW and around the NW section. Accordingly, to calculate the response functions (\ref{eq4}) we use a basis set of single-particle states restricted to the twelve states shown in Fig.~\ref{fig1}(c) to simplify our analysis. This clearly neglets higher energy excitations in the radial direction, since the QW higher states are not included. While we are not interested in quantitative predictions here, the excitations discussed in the following results are well converged with respect to the number of single-particle states except for the highest transitions.

The imaginary part of the response functions Eq.~(\ref{eq4}) have poles at energies corresponding to
electronic excitations in which the ground state is coupled with excited states through the density operator. We recall that within the
mean-field formalism employed here, the wave functions of the effective non-interacting system are single Slater determinants.
Hence, being the density operator a one-body operator, the only accessible excited states are described by Slater determinants differing in a single excitation from the determinant describing the ground state. All excitations are coupled by the dynamic Coulomb and exchange-correlation matrix in the response function. In the following we will adopt a widely used notation which labels the collective excitations with the single-particle transition of the final state with the largest contribution in the response function. We will also indicate the irrep of such final state as evaluated by multiplying the irreps of the two subbands participating in the single-particle transition.

\subsubsection{One occupied subband}
We first consider the case of the Fermi energy midway between subbands $n=1$ and $n=2, 3$, which corresponds to a linear electron
density of $\sim0.007\times10^7 \mathrm{cm^{-1}}$. The dispersion of CDEs calculated within the RPA is shown in Fig.~\ref{fig2}(a) 
as a function of $q_z/k_F$, with $k_F$ being the Fermi wave vector. One may recognize
\begin{itemize}
  \item one intra-subband CDE with vanishing energy at $q_z =0$;
  \item seven inter-subband CDEs associated with transitions from $n=1$ to each set of higher subbands, as indicated.
\end{itemize}
Since $n=1$ has irrep $\mathrm{A_{1g}}$, the symmetry of the excitation, which is the product of the irreps of the involved states, coincides with the irrep of the final state. Therefore, excitations to final states which are twofold degenerate are degenerate as well, and have, in general, larger spectral weight.

The dispersion of the inter-subband excitations in $q_z$ is characteristic of 1D electron systems.~\cite{TavaresPRB05}
At small $q_z$ the CDEs are blueshifted by the depolarization field from their analogous single-particle excitation, and approach the upper bound of single-particle continum as $q_z$ increases, without entering it and are not Landau damped.
The intra-subband CDE shows a typical dispersion for q1D systems~\cite{LiPRB91,DasSarmaPRB85,SatoPRB93,WendlerPRB94,WangPRA02}
proportional to $q_z\sqrt{|ln(q_z)|}$ in the long wavelength limit.

\begin{figure}[h!]
\includegraphics[width=0.8\textwidth]{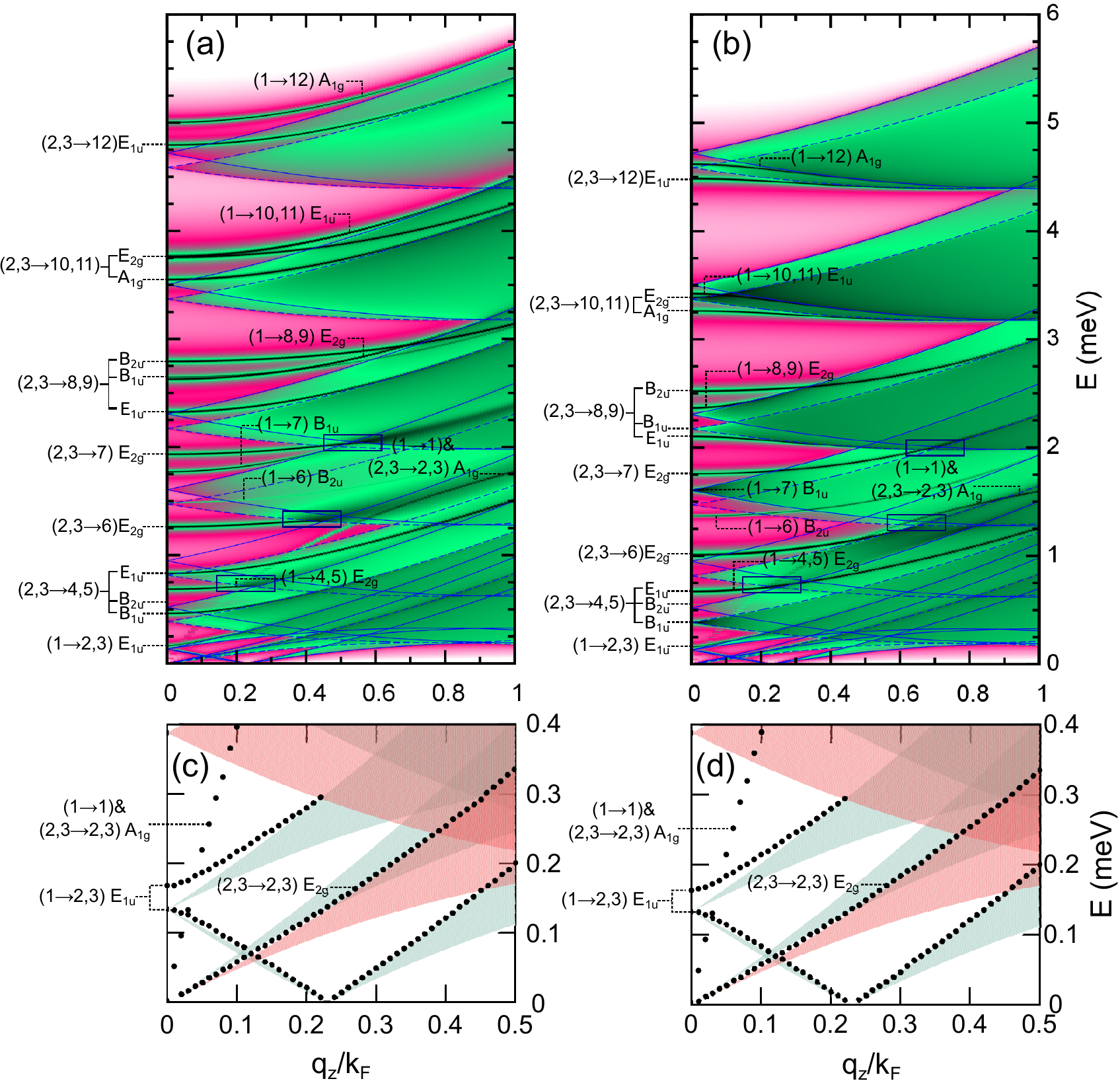}
\caption{CDEs dispersion with three occupied subbands calculated within the RPA (a) and the TDLDA  (b) approaches. The bounds
of the single-particle excitation continua for excitations from $n=1$ ($n=2,3$) to higher subbands are delimited by blue solid
(dashed) lines which can be calculated analytically from energy and momentum conservation. Dark-blue rectangles highlight
selected CDEs which are Landau damped. Panels (c) and (d) show low-energy CDEs between the occupied subbands calculated within the RPA and TDLDA, respectively. Black dots: CDEs; gray (red) area: single-particle continuum for transitions from $n=1$ ($n=2,3$).
The labels show the subbands involved in the excitation and the symmetry of the excited state.}
\label{fig3}
\end{figure}

\subsubsection{Three occupied subbands}

Next we consider a case with the Fermi energy midway between $n=2,3$ and $n=4,5$, with a linear electron density
$\sim0.039\times10^7 \mathrm{cm^{-1}}$. The RPA result is plotted in Fig~\ref{fig3}(a) and in an enlarged scale in Fig~\ref{fig3}(c). The low-energy CDEs consist in
\begin{itemize}
  \item two intra-subband CDEs corresponding to the occupied subbands;
  \item the same inter-subband CDEs from subband $n=1$ as in the single occupied subband case with an aditional CDE for the (1$\rightarrow$2,3) transition, with a negative dispersion in the long-wavelength limit, lying between the two branches of the (1$\rightarrow$2,3) single-particle continuum;
  \item multiple inter-subband transitions from $n=2,3$ to higher empty states.
\end{itemize}
Several additional issues may be discussed in this case.

\emph{CDEs degeneracies.} Since $n=2,3$ belong to the degenerate $E_{1u}$ set, inter-subband excitations to higher subbands lead to different number of CDEs, depending on the symmetry of the final state. Based on the product of the irreps of the involved subbands, we can identify the following cases:
\begin{itemize}
  \item if the final subband is non-degenerate, the CDE is twofold degenerate. For instance, (2,3$\rightarrow$7) yields the following degenerate irrep: $\mathrm{E_{1u}}\otimes\mathrm{B_{1u}}=\mathrm{E_{2g}}$.
  \item if the final subband is twofold degenerate and of the same symmetry as $n=2,3$, two CDEs appear, one being doubly degenerate; for instance, (2,3$\rightarrow$10,11) $\mathrm{E_{1u}}\otimes\mathrm{E_{1u}}= \mathrm{A_{1g}}\oplus\left[\mathrm{A_{2g}}\right]\oplus\mathrm{E_{2g}}$, where $\mathrm{A_{2g}}$ is basis of the antisymmetric representation of the permutation group.
  CDEs to such type of excited states are forbidden in the present spin-independent calculation since we always deal with singlet, and hence antisymmetric, spin wave functions which require symmetric orbital parts.
    \item if the final subband is twofold degenerate and it is not of the same symmetry as $n=2,3$, three CDEs arise,  one being doubly degenerate; for instance, for the excitation (2,3$\rightarrow$8,9) we have $\mathrm{E_{1u}}\otimes\mathrm{E_{2g}}=\mathrm{B_{1u}}\oplus\mathrm{B_{2u}} \oplus\mathrm{E_{1u}}$.
\end{itemize}

\emph{Depolarization shift.}
Inter-subband CDEs couple through off-diagonal elements of the dielectric tensor (see, e.g., Eqs.~(\ref{eq13}) and~(\ref{eq14})) only if they belong to the same symmetry. This leads to assorted depolarization shifts for different inter-subband CDEs in Fig.~\ref{fig3}(a). In general, however, these are larger than in Fig.~\ref{fig2}(a) due to the increase of electron density.

\begin{figure}[h!]
\includegraphics[width=0.5\textwidth]{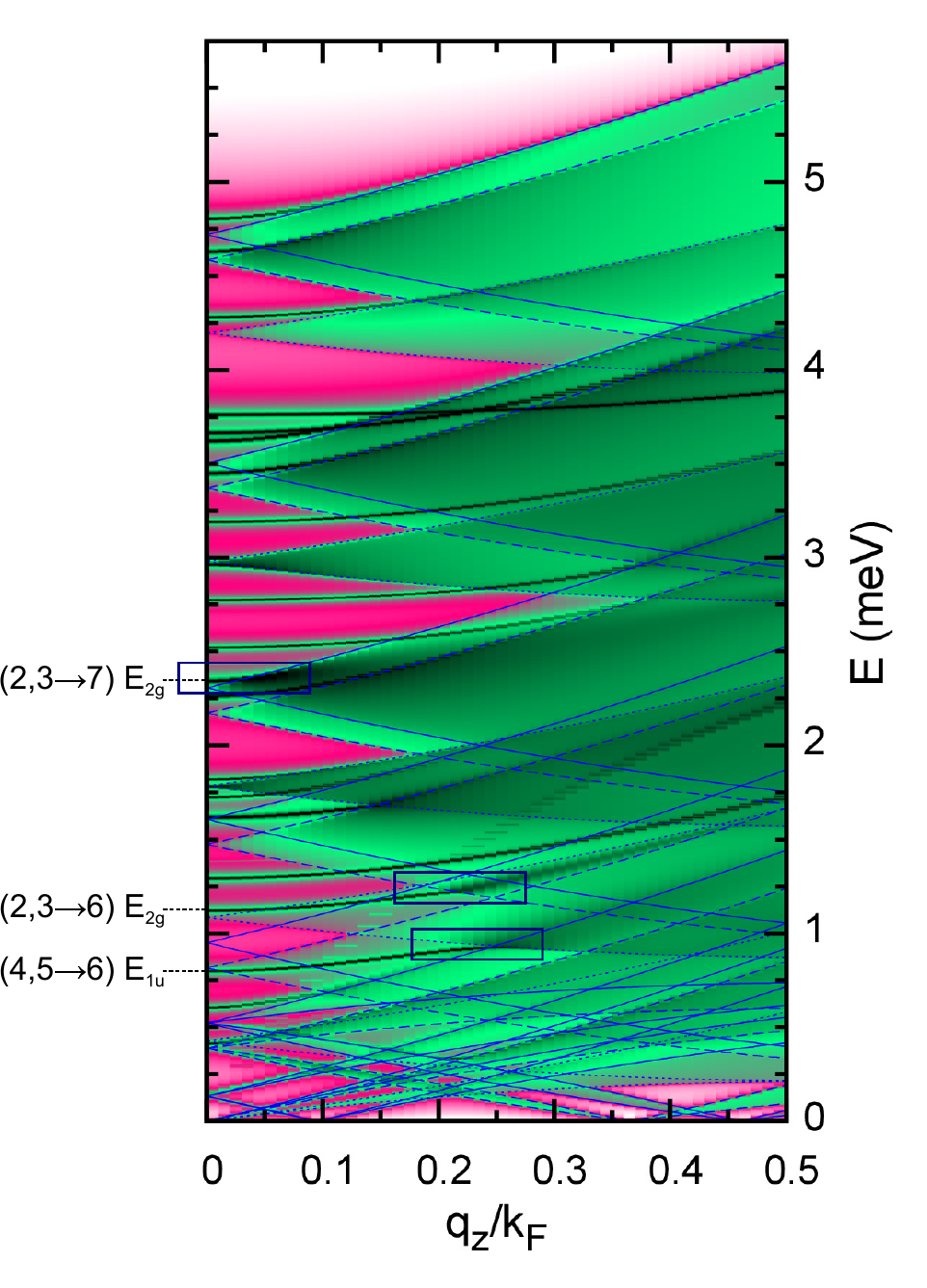}
\caption{CDEs dispersion for a NW with five occupied subbands calculated within the the TDLDA. The bounds
of the single-particle excitation continua from subband 1, (2,3), and (4,5) are delimited by blue solid, dashed and dotted lines, respectively,
and calculated analytically. The labels show the subbands involved in three selected excitations and the symmetry of the excited states. The dark-blue rectangles
illustrate selected Landau damping of these CDEs.}
\label{fig4}
\end{figure}

\emph{Landau damping.} Due to the large density of states, in this higher density case inter-subband CDEs often merge into
single-particle continuum and get Landau damped. However, \emph{plasmons are
Landau damped only when they enter the single-particle continua associated with transitions of the same symmetry}.
For example, plasmons (2,3$\rightarrow$7)$\mathrm{E_{2g}}$, (2,3$\rightarrow$6)$\mathrm{E_{2g}}$
and (1$\rightarrow$4,5)$\mathrm{E_{2g}}$ are  Landau damped when entering the lower bound of the single-particle continua
(1$\rightarrow$8,9)$\mathrm{E_{2g}}$, (2,3$\rightarrow$7)$\mathrm{E_{2g}}$ and (2,3$\rightarrow$6)$\mathrm{E_{2g}}$,
respectively, as shown in Fig.~\ref{fig3}(a).

\begin{figure}[h!]
\includegraphics[width=0.5\textwidth]{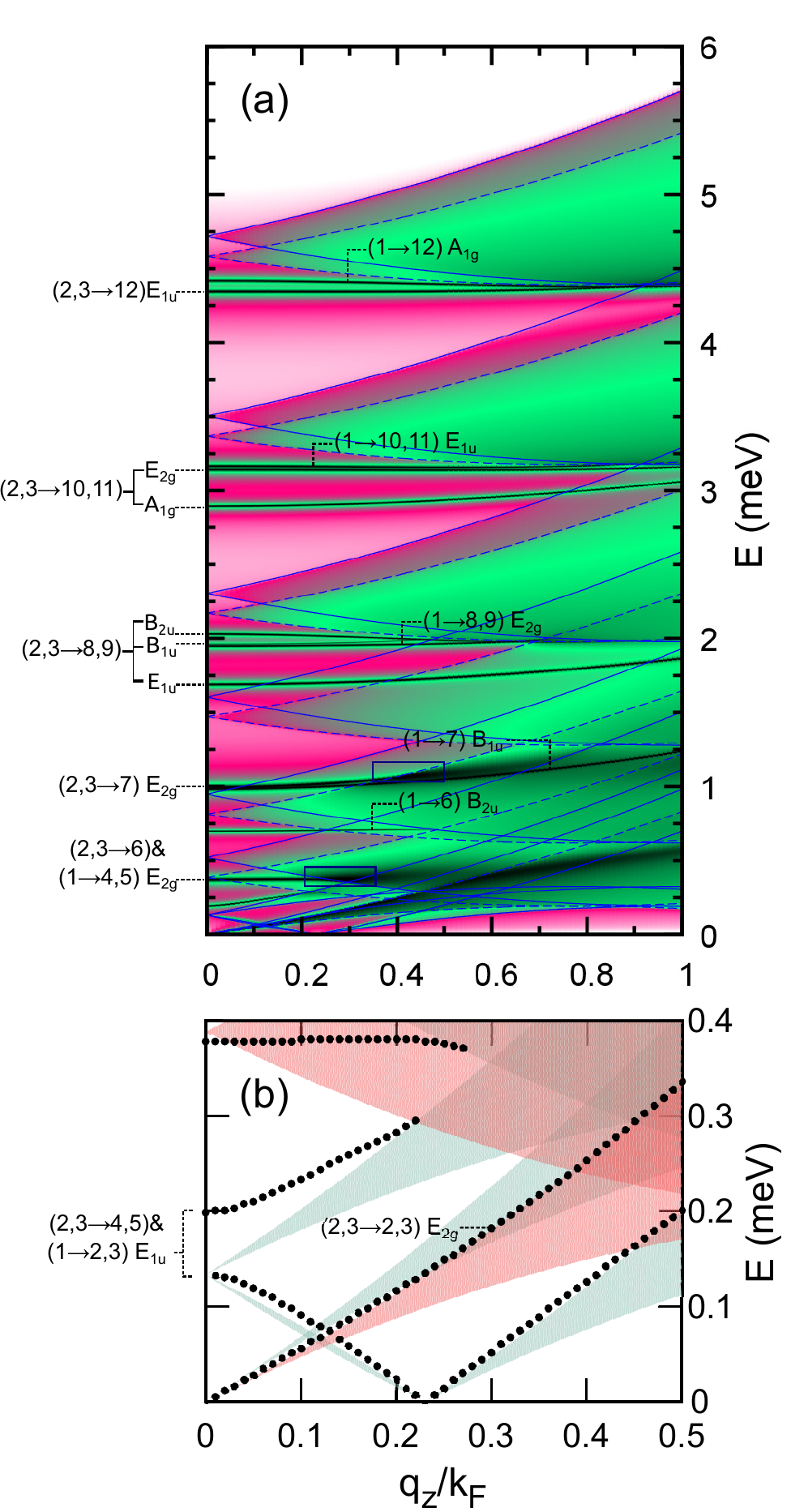}
\caption{(a) SDEs dispersion in $q_z/k_F$ for a NW with three occupied subbands calculated within the TDLDA. The bounds
of the single-particle excitation continua from subband 1, (2,3) are delimited by blue solid (dashed) lines calculated analytically.
The labels show the subbands involved in the excitations and the symmetry of the excited states. The dark-blue rectangles
illustrate selected Landau damping of SDEs. (b) Low-energy SDE spectrum for excitations between the occupied subbands calculated within the TDLDA.
The black dots show the SDEs dispersion, whereas the gray (red) area is the single-particle continuum for transitions from subband 1 (2,3).}
\label{fig5}
\end{figure}

\emph{Low-energy excitations.}
In Fig.~\ref{fig3}(c) we observe two inter-subband CDEs associated with the transition (1$\rightarrow$2,3). The one with low energy
shows an anomalous negative dispersion in the long wavelength limit. Such type of plasmon is exclusive from q1D systems with more than one occupied subband and is free of Landau damping for large ranges of $q_z$ as it lies between the bounds of the (1$\rightarrow$2,3) single-particle continuum.~\cite{MendozaPRB91,WendlerPRB91,WendlerPRB94}
The two intra-subband plasmons (1$\rightarrow$1) and (2,3$\rightarrow$2,3) are highlighted in Fig.~\ref{fig3}(c).
The (1$\rightarrow$1)$\mathrm{A_{1g}}$ excitation is more dispersive than the corresponding excitation in Fig.~\ref{fig2}(b).
This is due to the higher density regime and the coupling with the (2,3$\rightarrow$2,3) intra-subband plasmon.
The two CDEs couple because (2,3$\rightarrow$2,3) $\mathrm{E_{1u}}\otimes\mathrm{E_{1u}}=\mathrm{A_{1g}}\oplus\left[\mathrm{A_{2g}}\right]\oplus
\mathrm{E_{2g}}$ has a symmetry component in common with the (1$\rightarrow$1)$\mathrm{A_{1g}}$ plasmon.
On the other hand, the (2,3$\rightarrow$2,3)$\mathrm{E_{2g}}$ component appears as a slender acoustic plasmon~\cite{LeeJPC83}
lying between the two intra-subband single-particle continua (see Fig.~\ref{fig3}(c)). This type of plasmon free of Landau
damping is exclusive of q1D systems with more than one occupied subband.~\cite{LeeJPC83,GoniPRL91,MendozaPRB91,WendlerPRB91,WendlerPRB94,WangPRA02} The antisymmetric component (2,3$\rightarrow$2,3)$\mathrm{A_{2g}}$ is dark in the charge density channel for the same aforementioned reasons.

\emph{Echange and correlation effects.}
Fig.~\ref{fig3}(b) shows the CDEs calculated within the TDLDA. These are in one-to-one correspondence with the RPA CDEs, but
redshifted by the dynamic exchange-correlation matrix elements. The exchange-correlation vertex correction may also overcome
the direct Hartree term, bringing the plasmon below the corresponding single-particle excitations, as predicted by Das Sarma
\emph{et al.}~\cite{DasSarmaPRB93} and later observed by Ernst \emph{et al.}~\cite{ErnsPRL94} in very dilute QWs.
The symmetry selective Landau damping phenomena observed in the RPA spectrum are also present within the TDLDA, marked by dark-blue
squares in panel (c), although they are not so well resolved as in Fig.~\ref{fig3}(a).

\subsubsection{Five occupied subbands}
In Fig.~\ref{fig4} we show the CDEs for a case with the Fermi energy midway between $n=4,5$ and $n=6$ with a linear
electron density $\sim0.089\times10^7 \mathrm{cm^{-1}}$. Calculations are shown for TDFDT only. In this higher density regime all the inter-subband
plasmons appear blueshifted from their corresponding single-particle excitations in spite of the exchange-correlation corrections.
The same symmetry arguments used above to assign the excitation spectra apply also in this case and in higher subband occupations.
We do not include labeling of all CDEs appearing in Fig.~\ref{fig4} for the sake of conciseness. Landau damping of selected CDEs is
marked with dark-blue rectangles for plasmons (2,3$\rightarrow$7$)\mathrm{E_{2g}}$, (2,3$\rightarrow$6)$\mathrm{E_{2g}}$ and (4,5$\rightarrow$6)$\mathrm{E_{1u}}$, which are clearly Landau damped in the single-particle continua with the same symmetry, (1$\rightarrow$8,9)$\mathrm{E_{2g}}$, (2,3$\rightarrow$7)$\mathrm{E_{2g}}$ and (4,5$\rightarrow$7)$\mathrm{E_{1u}}$, respectively.

\subsubsection{Spin density excitations}

SDEs are computed from the imaginary part of the TDLDA reducible response function. In Fig.~\ref{fig5} we show SDEs for three occupied subbands.
SDEs appear in general redshifted with respect to their corresponding single-particle excitations. This is due to the so-called
excitonic shift caused by the exchange-correlation matrix elements. SDEs are less dispersive than CDEs (compare Figs.~\ref{fig5} and~\ref{fig3}).
This is originated from the fact that the inter-subband collective excitations become dispersive as they approach their corresponding
single-particle continua. However, the bottom bound of the continua is less dispersive than the top one. Hence, the SDEs merge in the single-particle continua at larger $q_z$. Once there the slope of the dispersion is also lower.

Additional differences with respect to CDEs are observed in the low-energy region:
\begin{itemize}
  \item less peaks are visible which, nevertheless, show higher configuration mixing. For example, at $\sim0.38$ and $\sim0.2$ meV there are two SDEs which have contributions from transitions (2,3$\rightarrow$6) plus (1$\rightarrow$4,5)$\mathrm{E_{2g}}$ and (2,3$\rightarrow$4,5) plus (1$\rightarrow$2,3)$\mathrm{E_{1u}}$, respectively;
  \item a single intra-subband SDE associated with transition (2,3$\rightarrow$2,3)$\mathrm{E_{2g}}$ is observed. It disperses almost linearly in $q_z$ between the two intra-subband single-particle continua as the slender acoustic plasmon observed in Fig.~\ref{fig3};
  \item SDEs also show Landau damping of inter-subband excitations. The most obvious are marked with dark-blue rectangles and correspond to the damping of SDEs (2,3$\rightarrow$7)$\mathrm{E_{2g}}$ and (2,3$\rightarrow$6) plus (1$\rightarrow$4,5)$\mathrm{E_{2g}}$ in the single-particle continua (2,3$\rightarrow$6)$\mathrm{E_{2g}}$ and (1$\rightarrow$4,5)$\mathrm{E_{2g}}$, respectively.
\end{itemize}

SDEs associated with transitions (2,3$\rightarrow$10,11) and (2,3$\rightarrow$2,3) lead to excited states of symmetry
$\mathrm{A_{1g}},\mathrm{E_{2g}}$ and $\mathrm{E_{2g}}$, respectively. Such excited states resulting from transitions between degenerate states of the same $\mathrm{E_{1u}}$ symmetry are a basis of the symmetric representation of the permutation group. 
Notice that here these are not proper excited states for SDEs, which implicate a triplet, and hence symmetric, spin function
requiring an antisymmetric orbital function. Therefore, for the two mentioned SDEs one would expect
to obtain the antisymmetric $[\mathrm{A_{2g}}]$ state. This physically incorrect result is a shortcoming consequence of the widely used spin-independent
formalism employed here.~\cite{KatayamaJPSJ85,LuoPRB93,MarmorkosPRB93}

\subsection{Inelastic light scattering spectra \label{ILS}}

We now focus on the ILS spectra in the non-resonant formalism discussed in section~\ref{theo_ILS} which has been widely used in spectroscopy of q2D and q1D electron systems (see, e.g., Refs.~\onlinecite{ILScardonaBook,SchullerBook2006}, and references therein). The authors have recently used this formalism to successfully assign ILS resoncances of a high-mobility EG in modulation-doped core-multishell NWs. \cite{FunkNL13}

\begin{figure}[h!]
\includegraphics[width=0.8\textwidth]{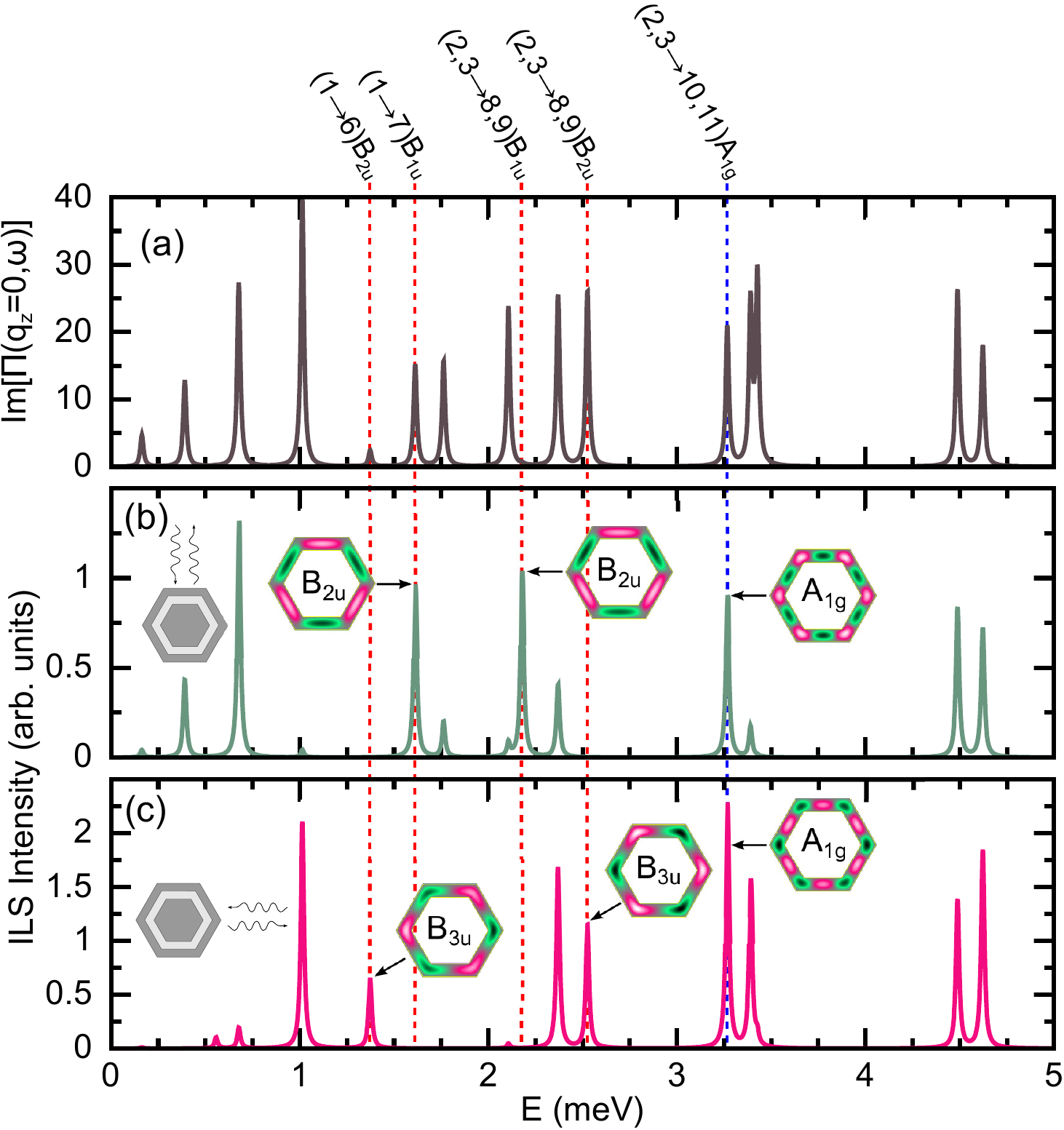}
\caption{(a) Imaginary part of the density-density response function Eq.~(\ref{eq4}) for a system with three occupied subbands and
$q_z=0$. (b) and (c) ILS spectra for the same system when photons are applied perpendicular to the top facet and along the maximal diameter, respectively, as illustrated in the left insets. The 2D colormap insets show the IDDs calculated at selected peaks and labeled with the excitation irreps of the $D_{2h}$ group. The vertical dashed lines are guides to the eye showing the energy position
of selected excitations indicated by the top labels.}
\label{fig6}
\end{figure}

The ILS cross section of CDEs is obtained from the imaginary part of the complete Fourier transform of the density-density response function, Eq.~(\ref{eq9}), which has a clear physical interpretation: the coefficients $\tilde{\Pi}_{ijlm}(q_z,\omega)$ give the spectral intensity of the CDEs or SDEs, while the matrix elements $\langle \phi_i(\mathbf{r})|e^{-i\,\mathbf{q}\mathbf{r}}|\phi_j(\mathbf{r}) \rangle \langle \phi_m(\mathbf{r}')|e^{i\,\mathbf{q}\mathbf{r'}} |\phi_l(\mathbf{r}') \rangle$ represent the coupling of light with electron charge density, which depend on the setup geometry. Therefore, the geometry of the experiment, defined by $\mathbf{Q}=\mathbf{Q}_i-\mathbf{Q}_s$ sets specific selection rules. This is well-known in q2D systems, where intra-subband or inter-subband excitations can be selectively excited by choosing the exchanged momentum along or perpendicular to the QW plane, respectively. In coQW the situation is clearly more complex. It is easy to realize that the photon momentum always has both in-plane and vertical components with respect to some of the facets, and one never recovers the ideal QW geometry.

Below, we will consider two backscattering configurations (see left insets in Fig.~\ref{fig6}), both with the incoming and scattered photons perpendicular to the NW axis:
\begin{itemize}
  \item[\emph{i)}] photons are perpendicular to the top/bottom NW facets;
  \item[\emph{ii)}] photons are parallel to the top/bottom NW facets;
\end{itemize}
Accordingly, we set $q_z=0$ and we assume a typical excitation energy of $1.92$ eV which yields $|\mathbf{Q}_i|=6.9\times10^5 \, \mathrm{cm^{-1}}$, and approximate $|\mathbf{Q}_s|=|\mathbf{Q}_i|$.

In the presence of the photons, the symmetry of the system is reduced from $\mathrm{D_{6h}}$ to $\mathrm{D_{2h}}$. In configuration \emph{i} the real and imaginary parts of the kernel, $e^{\pm i\,\mathbf{q}\mathbf{r}}$, have the irreps $\mathrm{A_{1g}}$ and $\mathrm{B_{2u}}$ of $\mathrm{D_{2h}}$ group, respectively: only those CDEs which have a $\mathrm{A_{1g}}$ or $\mathrm{B_{2u}}$ component in $\mathrm{D_{2h}}$ will be observed. Similarly, in configuration \emph{ii} the kernel have the irreps $\mathrm{A_{1g}}$ and $\mathrm{B_{3u}}$. This implies that some CDEs that are observed in one geometry, are dark in the other, and vice versa.

As an example, in Fig.~\ref{fig6}(b)(c) we show the calculated ILS spectra for a NW with three occupied subbands computed from the TDLDA density-density response function, with a damping parameter $\eta=0.01$ meV. In panel (a) we also show for comparison the imaginary part of the response function calculated at $q_z=0$, which has one peak for each CDE of the system. Peaks appear in both configurations if the corresponding excitations have irreps in $\mathrm{D_{6h}}$ which reduces to $\mathrm{A_{1g}}$ or to $\mathrm{B_{2u}}\oplus\mathrm{B_{3u}}$ in $\mathrm{D_{2h}}$. This is the case, for instance, of peak labelled (2,3$\rightarrow$10,11)$\mathrm{A_{1g}}$, which has a symmetric representation also under the effect of the photons, as it can also be observed in the IDDs shown in the insets. Some of the resonances, however, can only be observed in one of the scattering geometries. Peaks (1$\rightarrow$6)$\mathrm{B_{2u}}$ and (1$\rightarrow$7)$\mathrm{B_{1u}}$, for example, are only observed in configurations \emph{ii} and \emph{i} (panels (c) and (b)), respectively. Indeed, the irreps of these transitions in $\mathrm{D_{2h}}$ are $\mathrm{B_{3u}}$ and $\mathrm{B_{2u}}$, respectively. The corresponding IDD in the insets clearly shows the same symmetries. Due to the same argument, CDEs associated with the same transition involving degenerate subbands can selectively be observed in the different geometries. For instance, the peak (2,3$\rightarrow$8,9)$\mathrm{B_{1u}}$, which has a very low spectral weight (see panel (a)), is strongly enhanced in configuration \emph{i} (panel (b)) and dark in congiguration \emph{ii} (panel (c)). Peak (2,3$\rightarrow$8,9)$\mathrm{B_{2u}}$ shows the opposite behavior.

\section{Summary and conclusions \label{concl}}

We have used TDLDA and RPA methodologies to study electron collective excitations and ILS cross section of $\mathrm{GaAs/Al_{0.3}Ga_{0.7}As}$ core-multishell NWs. This is a complex system from the electronic point of view, where q1D and q2D channel coexist. The large dimensions of the target system have required a 3D computational scheme. We have shown how to make calculations manageable for such a complex geometry by exploiting the symmetries of the system and, particularly, we have developed a fast and accurate approach to calculate Coulomb matrix elements in hexagonal grids. The latter are shown to be necessary to obtain convergence and correct degeneracies with a reasonable grid density (details in Appendix \ref{coulomb_app}).

We have studied CDEs and SDEs of the EG at different density regimes, i.e., different number of occupied subbands of different localization (bents or facets) and degeneracy. We have identified a number of features ensuing from the discrete, here $\mathrm{D_{6h}}$, symmetry of the system:
\begin{itemize}
  \item inter-subband collective excitations, both CDEs and SDEs associated with transitions between twofold degenerate subbands split in different peaks in the spectra. The number of peaks is in agreement with the number of accessible excited states predicted by the symmetry group theory;
  \item we have observed symmetry selective Landau damping, namely, collective excitations are only Landau damped in single-particle continua associated with transitions of the same symmetry as the collective mode.
  \item we have observed intra-subband slender acoustic plasmons and inter-subband plasmons with negative dispersion exclusive of q1D systems with multiple subband occupation.
\end{itemize}

We have calculated the ILS spectra for two relevant scattering geometries in a backscattering configuration. As a result of the discrete symmetry of the system, the spectra are substantially anisotropic as the photon momentum is rotated around the NW axis. Some of the collective modes can be only observed in one of the geometries, being dark in the other. Selection rules are shown which explain the observed features. Although not included in the present study, ILS experiments may access excitations at larger energies where inter-subband excitations between excited QW states, i.e., with a radial nodal plane, are involved. Simulations in this higher energy range must also include coupling with GaAs phonon resonances.~\cite{FunkNL13}

\begin{acknowledgments}
We acknowledge partial financial support from APOSTD/2013/052 Generalitat Valenciana Vali+d Grant, EU-FP7 Initial Training Network INDEX, and University of Modena and Reggio emilia through grant 'Nano- and emerging materials and systems for sustainable technologies'
\end{acknowledgments}

\appendix*
\section{Coulomb matrix elements calculation in hexagonal grids\label{coulomb_app}}

The Coulomb matrix elements that we have to calculate are,

\begin{equation}
v_{ijkn}(q_z)= \int d\boldsymbol{r}
\int d\boldsymbol{r}' \, \phi_i(\boldsymbol{r})\, \phi_j^*(\boldsymbol{r}) \,
\hat{V}_C(\boldsymbol{r}-\boldsymbol{r}',q_z)\,
\phi^*_k(\boldsymbol{r}')\, \phi_n(\boldsymbol{r}')\, .
\label{eq_ap_1}
\end{equation}

\noindent These, by taking $g_{kn}(\mathbf{r}')=\phi^*_k(\boldsymbol{r}')\, \phi_n(\boldsymbol{r}')$ and
rearranging the integrals can be written as,

\begin{equation}
v_{ijkn}(q_z)= \int d\boldsymbol{r}\, \phi_i(\boldsymbol{r})\, \phi_j^*(\boldsymbol{r})
\int d\boldsymbol{r}' \, \hat{V}_C(\boldsymbol{r}-\boldsymbol{r}',q_z)\,g_{kn}(\mathbf{r}').
\label{eq_ap_2}
\end{equation}

\noindent In the above equation, $h(\mathbf{r})=\int d\boldsymbol{r}' \, \hat{V}_C(\boldsymbol{r}-\boldsymbol{r}',q_z)\,g_{kn}(\mathbf{r}')$
is the convolution of $\hat{V}_C(\boldsymbol{r}-\boldsymbol{r}',q_z)$ and $g_{kn}(\mathbf{r}')$. Therefore,
according to the Fourier convolution theorem, the Fourier transform of $h(\mathbf{r})$ can be obtained
as $\tilde{h}(\mathbf{q})=\tilde{V}_C(\mathbf{q},q_z)\,\tilde{g}_{kn}(\mathbf{q})$, where the tilde means a
Fourier transformed function in momentum space. Since $h(\mathbf{r})$ can be equivalently
obtained by performing the inverse Fourier transform of $\tilde{h}(\mathbf{q})$, i.e., $h(\mathbf{r})=
\mathcal{F}^{-1}\left[ \tilde{h}(\mathbf{q}) \right]$, it is possible to calculate the Coulomb matrix elements as,

\begin{equation}
v_{ijkn}(q_z)= \int d\boldsymbol{r} \phi_i(\boldsymbol{r})\, \phi_j^*(\boldsymbol{r})
\mathcal{F}^{-1}\left[\tilde{V}_C(\mathbf{q},q_z)\,\tilde{g}_{kn}(\mathbf{q})\right].
\label{eq_ap_3}
\end{equation}

With this approach the dimensionality of the Coulomb integrals is reduced from 4D to 2D at the expense
of performing three discrete Fourier transforms (DFTs). The outcome in terms of computation time is highly favorable thanks to the
existing fast Fourier transform (FFT) algorithms. Besides, $\tilde{V}_C(\mathbf{q},q_z)$ needs to be only calculated
once and it has a well known analytical expression,

\begin{equation}
\tilde{V}_C(\mathbf{q},q_z)=\frac{1}{(2\pi)^{3/2}}\frac{e^2}{\varepsilon_{\infty}\left( \mathbf{q}^2 + q_z^2 + q_D^2\right)},
\label{eq_ap_4}
\end{equation}

\noindent where $\varepsilon_{\infty}$ is the high frequency dielectric constant and $q_D$ is the momentum derived from the
Debye length $\lambda_D$ as $q_D=\frac{1}{\lambda_D}$. For the calculations shown in the present study we have employed
the GaAs dielectric constant, $\varepsilon_{\infty}=10.86$, and a Debye lenght $\lambda_D=1\,\mu m$.

Available libraries with implemented FFT algorithms exclusively work with numerical sampling on rectangular grids.
We have used this type of algorithms to calculate our Coulomb matrix elements by beforehand extrapolating our functions
onto a rectangular grid. However, such a procedure leads to qualitative errors in the Coulomb integrals.  Such errors, which
do not vanish as the grid is made denser, are due to the inefficacy of rectangular sampling to capture the implicit
hexagonal symmetry of the present system. For instance, in the first column of table~\ref{tab_ap} we show two diagonal
Coulomb matrix elements between the ground and the first twofold degenerated excited states calculated with a FFT
algorithm in a rectangular grid. The two of them should have the same value since the degeneracy is induced by the $\mathrm{D_{6h}}$
symmetry, however we observe a discrepancy of $\sim8\,meV \cdotp nm$. There exist alternative algorithms
\cite{EhrhardtSP93} in which the input data is sampled in
a hexagonal grid but the output of the Fourier transform is sampled on rectangular grids. Their use do not overcomes
the discrepancy in the Coulomb integrals calculation, though, as can be seen in the second column of table~\ref{tab_ap}.
A proper calculation of the Coulomb integrals requires the use of an algorithm that performs the DFT completely in a hexagonal
grid. Here we have adapted to our interest the formalism reported by Russel M. Mersereau~\cite{MersereauPIEEE79} initially
devoted to signal processing.

\begin{table}
\begin{tabular*}{0.75\textwidth}{@{\extracolsep{\fill} }  c | c  c  c  }

        & Rect. grid     &    Hex. grid$\rightarrow$Rect. grid &   Hex. grid \\
  \hline
  \hline
  $v_{1212}(k_F)$ [meV$\cdotp$nm] & 46.491 & 48.402 & 42.941\\
  $v_{1313}(k_F)$ [meV$\cdotp$nm] & 54.382 & 55.206 & 42.941\\

  \hline
\end{tabular*}

\caption{Diagonal Coulomb matrix elements between the ground state and the twofold degenerate first excited states
calculated with DFT algorithms working in three different grids (see text).}
\label{tab_ap}
\end{table}

The formulas appearing in reference~\onlinecite{MersereauPIEEE79} to perform the hexagonal discrete Fourier
transform (HDFT) of a 2D function $g(n_1,n_2)$ defined in a hexagonal domain are reported here for completeness:

\begin{eqnarray}
\tilde{g}(k_1,k_2)&=&\sum_{n_1,n_2 \in R_H(N,N)} g(n_1,n_2) \,\,
exp \left\{ -i\left[ \frac{\pi}{3N}(2n_1-n_2)(2k_1-k_2)+\frac{\pi}{N}n_2k_2 \right] \right\}, \\
g(n_1,n_2)&=& \sum_{k_1,k_2 \in R_H(N,N)} \tilde{g}(k_1,k_2)\,\,
exp \left\{ i\left[ \frac{\pi}{3N}(2n_1-n_2)(2k_1-k_2)+\frac{\pi}{N}n_2k_2 \right] \right\}.
\label{eq_ap_5}
\end{eqnarray}

\noindent Here, $n_{1,2}$ ($k_{1,2}$) are integer coordinates denoting points of the grid in the position (momentum)
space, and the sum is restricted to those points inside the domain limiting the grid $R_H(N,N)$. The choice of the
latter is not trivial since a DFT assumes that the input function is periodically replicated in the space. Moreover,
the replication pattern will determine the sampling of the output function of the DFT. Thus, in order to obtain $\tilde{g}(k_1,k_2)$
sampled on a hexagonal grid, one has to assume that $g(n_1,n_2)$ is periodically replicated with a hexagonal pattern.
This implies that the following relations have to be accomplished,

\begin{equation}
g(n_1,n_2)=g(n_1-3N,n_2)=g(n_1-2N,n_2-N)=g(n_1-N,n_2-2N),
\label{eq_ap_6}
\end{equation}

\noindent with N being the replication period.

\begin{figure}
\includegraphics[width=0.8\textwidth]{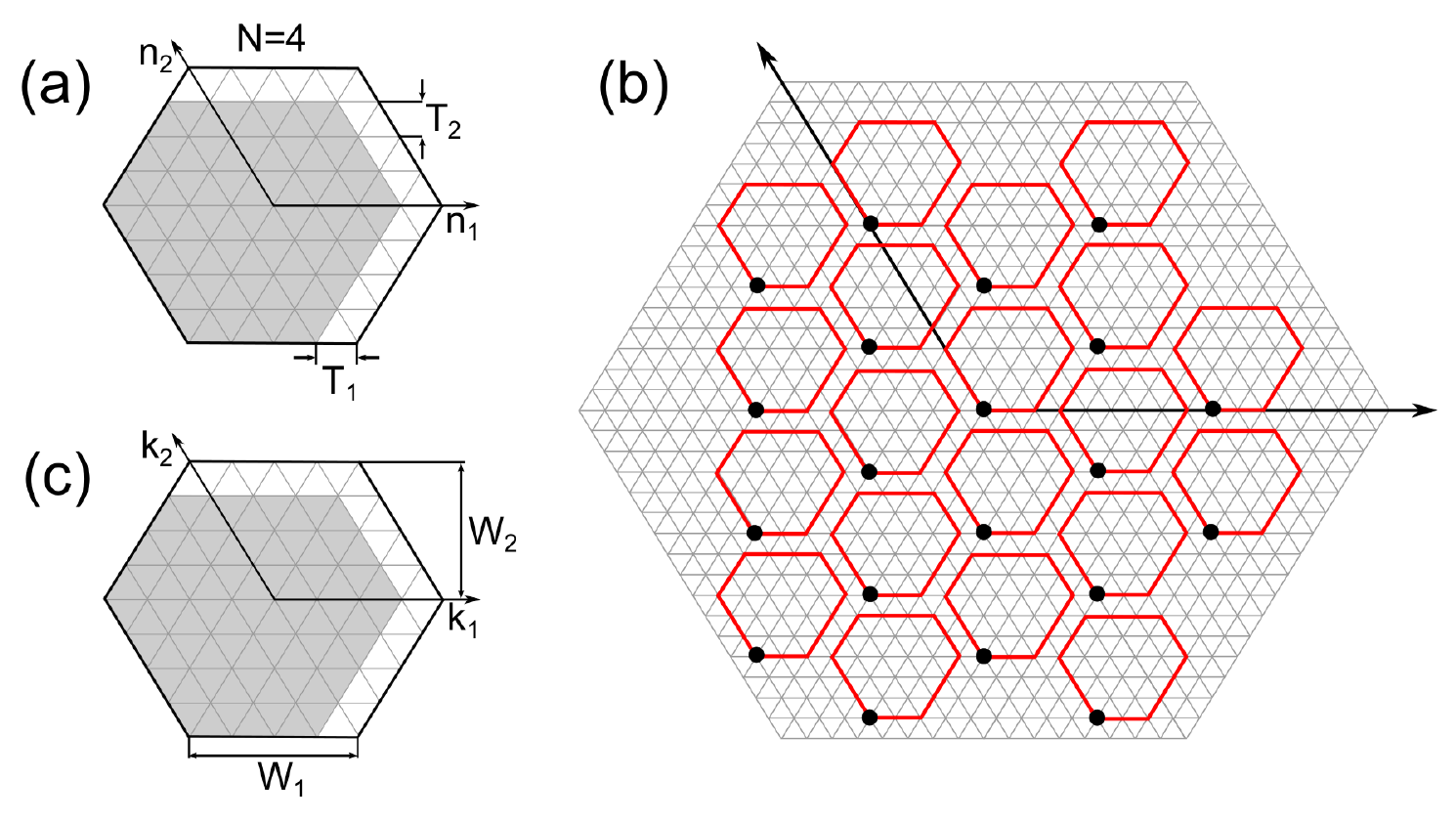}
\caption{(a)Schematic position space domain, (b) periodic replication with hexagonal pattern of the domain used for
the HDFT, and (c) schematic momentum space domain.}
\label{apen_fig}
\end{figure}

A regular hexagonal domain as the one shown in Fig.~\ref{apen_fig}(a) delimited by the solid line, which is equivalent to
the one in which our functions are sampled, is not a proper domain to perform the HDFT. The reason is that its
periodic replication according to Eqs.~(\ref{eq_ap_6}) leads to empty gaps and overlapping between the replicas,
which is known to produce aliasing. Instead, we use as $R_H(N,N)$ the deformed hexagonal region depicted by the
gray area in Fig.~\ref{apen_fig}(a), which can be exactly hexagonally replicated as illustrated in Fig.~\ref{apen_fig}(b).
Proceeding in this way we obtain an output function $\tilde{g}(k_1,k_2)$ hexagonally sampled on a equivalent domain
as the one illustrated by the gray area in Fig.~\ref{apen_fig}(c). The dimensions of the domain in the momentum
space, shown as $W_1$ and $W_2$ in Fig.~\ref{apen_fig}(c), are fixed by the sampling theorem as $W_1=\frac{4\pi}{3T_1}$
and $W_2=\frac{\pi}{T_2}$, where $T_1$ and $T_2$ are the sampling intervals in the position space (see Fig.~\ref{apen_fig}(a)).

As shown in the third column of Table~\ref{tab_ap} the use of the HDFT solves the symmetry induced numerical discrepancies
in the calculation of the Coulomb matrix elements. Its formal computation, though, is computationally more demanding
than in a rectangular sampling formalism, mainly due to the impossibility to separate the Fourier kernel in 1D-like DFTs. However,
there exist a fast Fourier implementation of the algorithm, fairly described in Ref.~\onlinecite{MersereauPIEEE79}, that will
allow one to increase the efficiency of the computation.

\bibliography{Elementary_excitations}

\end{document}